\documentclass{article}
\usepackage{graphicx}
\usepackage{amsmath,amssymb}
\usepackage{graphics,color,array,calc,rotating,epsfig,psfrag}
\numberwithin{equation}{section}
\usepackage{mathtools} 
\usepackage{amsmath}   
\usepackage{hyperref}
\usepackage{cite}
\usepackage{bm}
\usepackage{dcolumn}
\usepackage{float}
\usepackage{subfigure}
\usepackage{manyfoot}
\usepackage{epstopdf}
\usepackage{inputenc}
\usepackage{booktabs}  
\usepackage{siunitx}   
\usepackage{multirow}  
\DeclareNewFootnote{R}[roman]
\usepackage{amsfonts}
\usepackage[mathscr]{eucal}
\def\be{\begin{equation}} \def\ee{\end{equation}}
\def\bea{\begin{eqnarray}} \def\eea{\end{eqnarray}}

\newcommand{\nn}{\nonumber}
\begin{document}
\baselineskip 18pt%
\begin{titlepage}
\vspace*{1mm}%
\hfill%
\vspace*{15mm}%
\hfill
\vbox{
    \halign{#\hfil         \cr
          } 
      }  
\vspace*{20mm}

\begin{center}
{\large {\bf
Analytical Criteria for Potential-Well Formation in Non-Minimally Coupled Scalar-Tensor Theories: A Necessary Condition for Instability
}}\\
\vspace*{5mm}
{  Majid Karimabadi\footnoteR{ma.karimabadi@hsu.ac.ir}, Davood Mahdavian Yekta\footnoteR{d.mahdavian@hsu.ac.ir}, S. A. Alavi\footnoteR{s.alavi@hsu.ac.ir}}\\

\vspace*{0.2cm}
{$^{}$ Department of Physics, Hakim Sabzevari University, P.O. Box 397, Sabzevar, Iran}\\
\vspace*{1cm}
\end{center}
\begin{abstract}

Although the existence of a potential-well does not necessarily imply instability, its formation is a necessary prerequisite for the existence of unstable bound states. We investigate the formation of an effective potential-well for scalar-field perturbations in two non-minimally coupled scalar-tensor theories, involving couplings to the Ricci scalar and the Einstein tensor. In the context of the near-horizon approximation, we derive an analytical expression for the threshold coupling in each model at which the extremum of the effective potential reaches the event horizon. We use this criterion for the stationary, spherically symmetric black holes in this paper, in particular for a number of regular black holes spacetimes. We examine the dependence of the threshold couplings on the scalar field mass, angular multipole number, and the background parameters. At this distinguished coupling, the real part of the near-horizon quasi-normal mode frequencies vanishes, yielding purely imaginary frequencies. We further demonstrate that the universal area spectrum of spherically symmetric black holes can be recovered from the near-horizon frequency spectrum without invoking the highly damped-mode approximation. Remarkably, this spectrum is independent of the specific non-minimal coupling scheme.

\end{abstract}

\end{titlepage}

\section*{Introduction}

Studying the physical properties of black holes, as fascinating predictions of general relativity (GR), has attracted considerable attention since the first observations of gravitational lensing and, more recently, the direct detection of gravitational waves by the LIGO and Virgo collaborations \cite{LIGOScientific:2016aoc,LIGOScientific:2016sjg}, as well as the imaging of black hole shadows by the Event Horizon Telescope (EHT) \cite{Akiyama:2019cqa}. In particular, gravitational waves emitted during the final stages of binary black hole mergers have opened a new observational window into the strong-field regime of gravity. The final stage of such mergers, known as the ringdown phase, contains valuable information about a perturbed black hole as it relaxes toward a stationary configuration.

The relaxation process can be described in terms of quasi-normal modes (QNMs), which characterize the response of a black hole to perturbations induced by scalar, vector, or gravitational fields. These modes are characterized by complex frequencies whose real and imaginary parts describe the oscillation and damping behavior of the perturbations, respectively. The QNM spectrum therefore provides important information about the spacetime geometry and the dynamical stability of black holes. For comprehensive reviews of black hole QNMs, see, for example, Refs. \cite{Berti:2009kk,Konoplya:2011qq}.

The stability of black holes under perturbations provides an important test of GR and its extensions. In particular, the effective potential governing perturbations plays a crucial role in determining the dynamical behavior of the corresponding modes. The formation of a potential well outside the event horizon is an important prerequisite for the existence of an unstable bound state. Nevertheless, the mere existence of such a well does not necessarily imply instability, since its actual onset depends on the properties of the potential and the associated boundary conditions. Recent studies have also investigated purely imaginary QNMs and their connection with non-oscillatory, growing ringdown profiles \cite{Ma:2026eaf,Capuano:2026tjy}. These developments motivate a closer investigation of the effective potential structure and its near-horizon behavior in different gravitational theories.

In this work, we investigate the formation of effective potential wells for scalar-field perturbations in static, spherically symmetric black hole spacetimes within two non-minimally coupled scalar-tensor models. The first model describes a direct coupling between the scalar field and the Ricci scalar, whereas the second involves a derivative coupling to the Einstein tensor. Such non-minimal interactions modify the effective potential governing scalar perturbations and can lead to the emergence of a potential well outside the event horizon as the coupling strength increases. Although our analytical results are general for static, spherically symmetric black hole geometries, we consider regular black hole solutions as illustrative examples to examine the implications of these criteria. 

In classical GR, singularities inside the black holes are closely related to the gravitational collapse of matter \cite{Hawking:1965mf,Penrose:1964wq}. Resolving these singularities, the so-called regular black holes (RBHs), remains one of the fundamental challenges in theoretical physics and have attracted considerable interest. Their construction generally requires either suitable modifications of the gravitational theory or the introduction of nontrivial matter sources. The pioneering approaches of Bardeen \cite{Bardeen} and Borde \cite{Borde:1994ai,Borde:1996df} introduced RBH geometries through suitable modifications of the mass function. Another important approach involves embedding RBH solutions into nonlinear electrodynamics (NLED) \cite{AyonBeato:1998ub,AyonBeato:1999ec,AyonBeato:2000zs,Bronnikov:2000yz,Dymnikova:2004zc,Hayward:2005gi}, leading to several interesting generalizations of charged black holes. Alternatively, noncommutative (NC) geometry provides a framework for modifying classical black hole solutions. In particular, the coordinate coherent state formalism replaces the point-like mass and charge distributions with smeared Gaussian profiles, thereby yielding regular spacetime geometries \cite{Nicolini:2005vd,Ansoldi:2006vg,Nicolini:2008aj}.

Our main objective is to derive analytical expressions for the non-minimal couplings at which an extremum of the effective potential reaches the event horizon. Using the near-horizon approximation, we show that these expressions determine the threshold for the formation of an effective potential well in the exterior region. This threshold characterizes a stage preceding the onset of instability and should not be identified with the critical coupling at which an unstable mode first appears. Our analytical criteria also reveal how the onset of potential-well formation depends on the scalar-field mass, angular multipole number, and parameters characterizing the background geometry. The significance of the near-horizon regime extends beyond the present analysis, with applications to quantum critical phenomena near black hole horizons \cite{Horvath:2019xcc}, universal scaling laws in gravitational collapse \cite{Gralla:2018xzo}, holographic dualities of near-horizon geometries \cite{Hadar:2022xag,Lewandowski:2017wfa}, and the study of QNMs in the near-horizon region \cite{Das:2024fwg,MahdavianYekta:2019pol}.

We further investigate the QNM spectrum in the near-horizon region at the threshold couplings corresponding to the onset of potential-well formation. In particular, we show that the real part of the corresponding QNM frequencies vanishes, which means that the modes include only the purely imaginary frequencies, a result that has been considered in recent literature \cite{Ma:2026eaf,Capuano:2026tjy}. Moreover, we demonstrate that the universal area spectrum of spherically symmetric black holes can be recovered directly from the near-horizon frequency spectrum at this coupling, without invoking the highly damped-mode approximation discussed in Refs.~\cite{Maggiore:2007nq,Vagenas:2008yi}. Remarkably, the resulting area spectrum is independent of the specific non-minimal coupling scheme.

The paper is organized as follows. In Sec. \ref{sec1}, we review two classes of RBH geometries that serve as illustrative examples in our analysis. The first includes the NC Schwarzschild spacetime in the coordinate coherent state formalism, while the second comprises black holes supported by NLED, including the Bardeen, Hayward, and Ayón-Beato-García (ABG) solutions. In Sec. \ref{sec2}, we consider the two non-minimally coupled scalar-field models and derive their corresponding perturbation equations. In Sec. \ref{sec3}, we investigate the formation of the effective potential well and derive general analytical expressions for the corresponding threshold couplings using the near-horizon approximation. We also examine the associated QNM spectrum and its implications for black hole area quantization. We will provide a breif summary and discussion of our results in Sec. \ref{sec4}. 

\section{Regular black holes} \label{sec1}

 The spacetimes free of curvature singularities that are alternatives to singular mathematical black holes predicted by GR \cite{Hawking:1965mf} have received
much attention in recent years as viable effective descriptions of quantum corrected
gravitational collapse \cite{Penrose:1964wq}. In most of the constructions, this regularity is achieved either by adding an effective matter sources like NLED or by considering the NC effects of the spacetime while preserving the existence of an event horizon. In the following section we will review the structures of RBHs belonging to both of these approaches.

\subsection{NC geometry in coordinate coherent state formalism}

The incorporation of NC geometry into the gravitational theories is given by the standard form of the Einstein tensor in GR accompanied with a modified energy-momentum tensor as a source on the right hand side \cite{Smailagic:2003yb}, which is a gravitational analogue of the NC modification of quantum field theory. One of the most widely used approaches to implement noncommutativity is the \textit{coordinate coherent state formalism} (for more about this formalism one can see the review \cite{Nicolini:2008aj}). In this framework, the effects of noncommutativity are implemented by replacing the classical point-like mass and charge distributions with smeared Gaussian profiles \cite{Nicolini:2005vd,Ansoldi:2006vg}, there by avoiding the curvature singularity  of classical solutions. This method does not deform the spacetime manifold itself, but rather modifies the matter source while preserving the general covariance. We have done the study of gravitational measurements in four and higher dimensions and the QNM spectra of these RBHs in \cite{Karimabadi:2018sev,Yekta:2019wlw,MahdavianYekta:2019pol}

Solving the Einstein field equations with a mass density of spherically symmetric particle-like gravitational source
\be \label{md}
\rho_{\theta}=\frac{M}{(4\pi\theta)^{3/2}\exp(-r^2/4\theta)},
\ee
 where $\theta$ is the NC parameter of dimension length squared, we can find the standard Schwarzschild black hole with NC correction that regularizes the curvature at short distances \cite{Nicolini:2005vd}, but it remains spherically symmetric and static as follows
\be\label{bm}
ds^2 = -f(r)\,dt^2 + \frac{dr^2}{f(r)} + r^2 \left(d\vartheta^2 + \sin^2\vartheta\,d\varphi^2\right),
\ee
where the blackening function is given by 
\be\label{Sch}
f(r) = 1 - \frac{4M}{r\sqrt{\pi}} \gamma\left(\frac{3}{2}, \frac{r^2}{4\theta}\right).
\ee
$\gamma(a, x)$ is the incomplete Gamma function 
\be
\gamma\left(\frac{3}{2}, \frac{r^2}{4\theta}\right)=\frac{\sqrt{\pi}}{2}-\Gamma\left(\frac{3}{2}, \frac{r^2}{4\theta}\right).
\ee
It can be mathematically checked that the smeared mass distribution (\ref{md}) removes the curvature singularity at $r = 0$, i.e., the curvature invariants are finite there \cite{MahdavianYekta:2019pol,Karimabadi:2025gpk}. In addition, for such RBH, it is shown that the condition for the existence of horizon(s) is \cite{ Karimabadi:2018sev}
\be \label{ran}
\theta \leq \left(\frac{M}{1.9}\right)^2.
\ee


\subsection{Non-linear electrodynamic source}

As a generalization of usual Maxwell theory, the NLED was proposed to remove the divergence of self-energy of the electromagnetic field of a point charge (for an introductory notes on NLED see e.g. \cite{Sorokin:2021tge}). It has also been shown that NLED can be a material source of gravity to obtain non-singular RBHs in GR and various alternative theories. In general, for all models in GR a non-linear Lagrangian of Maxwell electrodynamics is coupled to Einstein gravity just like $\mathcal{L} (f,h)$ where $f=\frac14 F_{\mu\nu} F^{\mu\nu}$  and $h=\frac14\, ^{*} F_{\mu\nu} F^{\mu\nu}$. Here, $F_{\mu\nu}=2\nabla_{[\mu}A_{\nu]}$ is the field strength and $^{*} F_{\mu\nu}=\frac12 \sqrt{-g}\, \varepsilon_{\mu\nu\rho\sigma}  F^{\rho\sigma}$ is its Hodge dual. 

Following Refs. \cite{AyonBeato:1998ub,AyonBeato:2000zs,AyonBeato:1999ec}, the Lagrangian that will be considered in this paper only includes the field $f$, i.e., the action is given as follows 
\be \label{act} S=\int d^4 x\,\sqrt{-g}\left(\frac{1}{16\pi}\,R-\frac{1}{4\pi} \,\mathcal{L} (F)\right),\ee
where $R$ is the Ricci scalar and we set $G=1$ for later convenience. The NLED source is described by
\be\label{NLE}\mathcal{L} (f)=\frac{3}{2\alpha q^2}\left(\frac{\sqrt{2q^2 f}}{1+\sqrt{2q^2 f}}\right)^{5/2},\ee
here the parameter $\alpha$ is related to the charge and mass of the black hole as $\alpha\equiv \frac{|q|}{2M}$. Though the Bardeen black hole as proposed initially, was not an exact solution to the Einstein equations and the quantity $q$ was left as a regularizing parameter of the theory without any physical interpretation, but later ABG had provided RBHs with physical interpretation for $q$ as the monopole charge of a self-gravitating magnetic field of NLED\cite{AyonBeato:1998ub}. The line element of static, spherically symmetric solution for the field equations of action (\ref{act}) can also be described by (\ref{bm}) as well, with the blackening factor 
\be \label{func} f(r)=1-\frac{2 m(r)}{r}\,.\ee

In this paper we investigate the instability of field perturbations for three families of RBHs in NLED scheme with different mass functions;
\begin{itemize}
\item Bardeen RBH,
\be \label{Bardeen} m(r)=\frac{Mr^3}{(r^2+q^2)^{3/2}}\,,\ee
\item Hayward RBH,
\be \label{Hayward} m(r)=\frac{Mr^3}{(r^3+q^3)}\,,\ee
\item ABG RBH,
\be \label{ABG} m(r)=\frac{Mr^3}{(r^2+q^2)^{3/2}}-\frac{q^2 r^3}{2(r^2+q^2)^2}.\ee
\end{itemize}
In the asymptotic limit $r\rightarrow \infty$, the solutions (\ref{Bardeen}) and (\ref{Hayward}) behave as an asymptotic flat Schwarzschild black hole with mass $M$, while the third one (\ref{ABG}) behaves as a Reissner-Nordstr\"{o}m black hole with mass $M$ and charge $q$. On the other side, in the limit $r\rightarrow 0$ each solution behaves as a de Sitter spacetime \cite{Toshmatov:2015wga}.
The regularity of these solutions can also be verified by computing the curvature invariants in this limit \cite{MahdavianYekta:2019pol}.
In addition,  depending on the value of $q$, they may have no horizon at all, one degenerate horizon, or two distinct horizons. For $M=1$, the extremal values of the charge for Bardeen, Hayward, and ABG RBHs are respectively given by $q_{B}\sim0.7698$,  $q_{H}\sim1.0582$, and  $q_{A}\sim0.6342$ \cite{MahdavianYekta:2019pol}.

\section{Non-minimally coupled models}\label{sec2}

The most general extension of GR based on the inclusion of a scalar degree of freedom, the so-called scalar-tensor theory, is the Horndeski gravity \cite{Horndeski:1974wa}. However, the general structure of Horndeski gravity is quite complicated which is the main obstacle preventing its application for most general setup for solution and analysis of some particular problems in cosmology and GR. Therefore, some particular simple cases of general Horndeski gravity are usually used, the theories with non-minimal coupling between gravitational and scalar degrees of freedom. In this section we consider the dynamics of a massive scalar field which is non-minimally coupled under two distinct schemes. In the first the scalar field is coupled directly to the Ricci scalar of the RBH geometries, while in the latter its derivatives are coupled to the components of the Einstein tensor. These couplings are extensively used in the cosmological models and effective field theories.   

\subsection{Scalar coupling model}

Suppose that a real field $\phi$ of mass $\mu$ is non-minimally coupled to the Ricci scalar $R$ of the background geometry. This theory is described by the following action \cite{Clifton:2011jh}
\be \label{sact}
S = -\frac{1}{2} \int d^4x\, \sqrt{-g} \left( g^{\mu\nu} \nabla_\mu \phi \nabla_\nu \phi + \mu^2 \phi^2 + \zeta R \phi^2 \right),
\ee
where $\zeta$ is a dimensionless coupling constant. Varying the action with respect to $\phi$ yields a modified Klein-Gordon equation as
\be \label{eomsc}
\left[ \Box - \mu^2 - \zeta R \right] \phi(t, r, \vartheta, \varphi) = 0,
\ee
where $\Box \equiv \nabla_\mu \nabla^\mu$. This equation shows that this kind of coupling plays the role of a mass term for constant curvature backgrounds. In order to study the perturbations of this field in the next sections we choose the following ansatz with usual decomposition in spherical coordinates
\be \label{ansatz}
\phi(t, r, \vartheta, \varphi) = A(r) Y_{\ell m}(\vartheta, \varphi) e^{-i \omega t},
\ee
where $Y_{\ell m}$ are spherical harmonics. Introducing the redefinition $A(r) = \psi(r)/r$ \footnote{This redefinition of radial function in d-dimensional spacetime is given by $A(r) =\frac{ \psi(r)}{r^{(d-2)/2}}$, of course in the case of Ricci-coupled model.} transforms the master equation (\ref{eomsc}) into a Schr\"{o}dinger-like form
\be \label{req1}
- f(r) \frac{d}{dr} \left[ f(r) \frac{d}{dr} \psi(r) \right] + V(r) \psi(r) = \omega^2 \psi(r).
\ee
Using the tortoise coordinate $dr^*/dr = 1/f(r)$, one obtains the Regge-Wheeler equation \cite{Regge:1957td}:
\be \label{sceq}
\frac{d^2 \psi}{d{r^*}^2} + \left[ \omega^2 - V(r) \right] \psi = 0,
\ee
with the effective potential
\be \label{veff}
V(r) = f(r) \left( \mu^2 + \frac{\ell(\ell+1)}{r^2} + \frac{f'(r)}{r} + \zeta R \right).
\ee
\subsection{Tensor coupling model}

Now consider again a massive scalar field which its derivatives are non-minimally coupled to the Einstein tensor $G^{\mu\nu}$. In other words, the scalar field would be kinetically coupled with the spacetime geometry of a RBH. The describing action is given by \cite{Sushkov:2009hk}
\begin{equation}\label{ss}
S = -\frac{1}{2} \int d^4x\, \sqrt{-g} \left( g^{\mu\nu} \partial_\mu \phi \partial_\nu \phi + \mu^2 \phi^2 - \hat{\zeta} G^{\mu\nu} \partial_\mu \phi \partial_\nu \phi \right),
\end{equation}
where $\hat{\zeta}$ is a coupling constant of length squared dimension in the geometric units where \(M\) has the dimension of length, such that \(\hat\zeta\) can be defined in terms of \(\zeta\) as \(\hat\zeta = M^2 \zeta\). The scalar field equation is obtain by varying this action with respect to $\phi$ as follows
\begin{equation}\label{Eom1}
\frac{1}{\sqrt{-g}} \partial_{\mu} \left[ \sqrt{-g} \left( g^{\mu \nu }- \hat\zeta G^{\mu \nu} \right) \partial _{\nu} \phi \right] - \mu ^{2} \phi = 0.
\end{equation}

Using the ansatz (\ref{ansatz}) by replacing $A(r)$ with $B(r)$ leads to the following second-order radial differential equation 
\begin{equation} \label{req2}
\frac{1}{r^2} \frac{d}{dr} \left[ r^2 f(r) \bigl(1 + \hat\zeta K(r) \bigr) B'(r) \right] +\frac{\omega^2\bigl(1 + \hat\zeta K(r) \bigr)}{f(r)} B(r)-\left( \mu^2 + \frac{\ell(\ell+1)}{r^2} \bigl(1 - \hat\zeta L(r)\bigr) \right)B(r) = 0,
\end{equation}
which is similar to Eq.~(\ref{req1}) and 
\begin{equation} \label{kl}
K(r) = \frac{1 - f(r)}{r^2} - \frac{f'(r)}{r}, \qquad L(r) = K(r) - \frac{1}{2} R,
\end{equation}
where prime is the derivative with respect to $r$ and $R$ is the Ricci scalar of the background geometry. This equation can be recast for convenience in the following form 
\begin{equation} \label{E1}
f(r)^2 B''(r) + \Sigma(r) B'(r) +\left(\omega^2+ \eta(r)\right) B(r) = 0,
\end{equation} 
where
\begin{align}
\Sigma(r) &= f(r)^2 \left( \frac{\hat\zeta K'(r)}{1+\hat\zeta K(r)} + \frac{f'(r)}{f(r)} + \frac{2}{r} \right), \label{H1} \\
\eta(r) &= \frac{-f(r)}{1+\hat\zeta K(r)} \left( \frac{\ell(\ell+1)}{r^2}\bigl(1 - \hat\zeta L(r)\bigr) + \mu^2 \right). 
\end{align}
After the field redefinition
\begin{equation} \label{Xi}
B(r) = \frac{1}{r \left(1 + \hat\zeta K(r) \right)^{1/2}} \psi(r),
\end{equation}
and introducing the tortoise coordinate $r^*$ as done in the previous subsection, and doing some lengthy calculation we obtain a Regge-Wheeler equation (\ref{sceq}) with the effective potential 
\begin{equation} \label{VT}
V(r) = \frac{f(r) f'(r)}{r} + f(r)^2 H(r) - \eta(r).
\end{equation}
and the function $H(r)$ is defined by
\begin{equation}
H(r) = \frac{\hat\zeta}{1+\hat\zeta K(r)} \left( \frac{K''(r)}{2} + \frac{K'(r) f'(r)}{2f(r)} + \frac{K'(r)}{r} \right) - \frac{1}{4} \left( \frac{\hat\zeta K'(r)}{1+\hat\zeta K(r)} \right)^2. \label{H2}
\end{equation}
It is worth mentioning that though both couplings $\zeta$ and $\hat\zeta$ are dimensionally different in these schemes, but as previously denoted by \(\hat\zeta = M^2 \zeta\) they can be identified as the same coupling when one chooses \(M=1\) in the calculations. 
\section{Potential-well formation and instability} \label{sec3}

Recently, we have extensively studied the ringdown profiles of scalar-field perturbations in the vicinity of NC Schwarzschild black holes in both coupling models \cite{Karimabadi:2025gpk}, using the numerical time-domain integration method \cite{Gundlach:1993tn}. Our numerical investigations revealed that, for certain regions of the parameter space, the late-time behavior of the ringdown profiles can significantly deviate from the usual decaying power-law tail and may eventually exhibit an instability. In the present work, we focus on the formation of an effective potential well as an important prerequisite for the development of such instabilities. In particular, we identify a characteristic value of the coupling parameter, denoted by $\zeta_{\text{well}}$, at which the first extremum of the effective potential reaches the event horizon. We first provide a numerical and schematic description of the emergence of the potential well for NC Schwarzschild black holes and then, using the near-horizon approximation, derive general analytical expressions for $\zeta_{\text{well}}$ in static, spherically symmetric black hole backgrounds, including the RBH geometries introduced in Sec.~(\ref{sec2}).

\begin{figure}[H]
\centering
{\includegraphics[width=.49\textwidth]{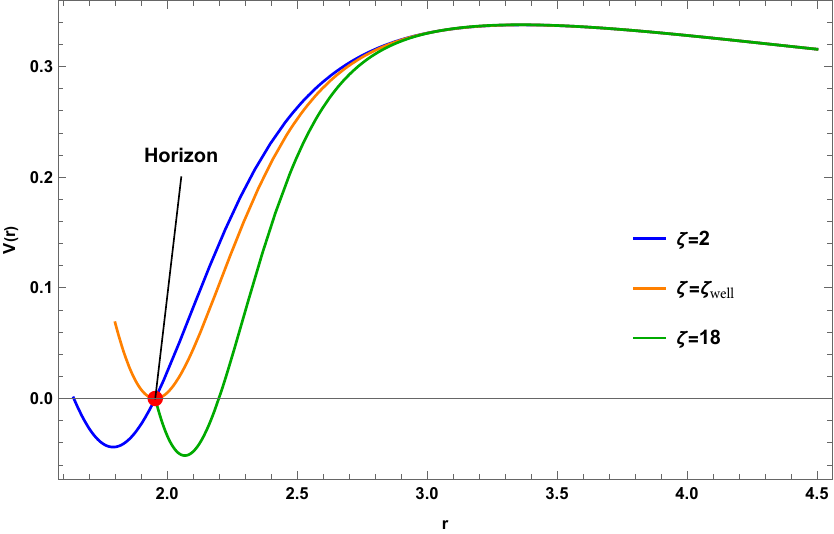}}
\caption{Effective potentials for NC Schwarzschild black hole when $\theta=0.2$, $\ell=2$, and $\mu=0.5$. The threshold value for the potential-well formation is $\zeta_{\text{well}}=5.673$.}
\label{f1}
\end{figure}

As discussed in Refs.~\cite{Karimabadi:2025gpk,MahdavianYekta:2019pol}, the effective potential in the Regge--Wheeler equation can develop a well just outside the event horizon for particular values of the coupling parameter $\zeta$. When the coupling increases beyond the corresponding threshold, the potential well becomes deeper. The formation of such a well is a necessary prerequisite for the existence of unstable bound states; however, its mere existence does not necessarily guarantee the onset of instability. As illustrated in Fig.~(\ref{f1}) for NC Schwarzschild black holes in the tensor-coupled model, when $\zeta=\zeta_{\text{well}}$, the first extremum of the effective potential is located precisely at the event horizon, $r=r_h$ (orange curve). For $\zeta<\zeta_{\text{well}}$, the corresponding well lies inside the event horizon (blue curve), whereas for $\zeta>\zeta_{\text{well}}$, the well emerges outside the horizon (green curve). Therefore, $\zeta_{\text{well}}$ characterizes the onset of effective potential-well formation rather than the actual threshold for instability. Note that, for all numerical calculations in this paper, we set the mass of the black hole to $M=1$.

\begin{figure}[H]
\centering
{\includegraphics[width=.50\textwidth]{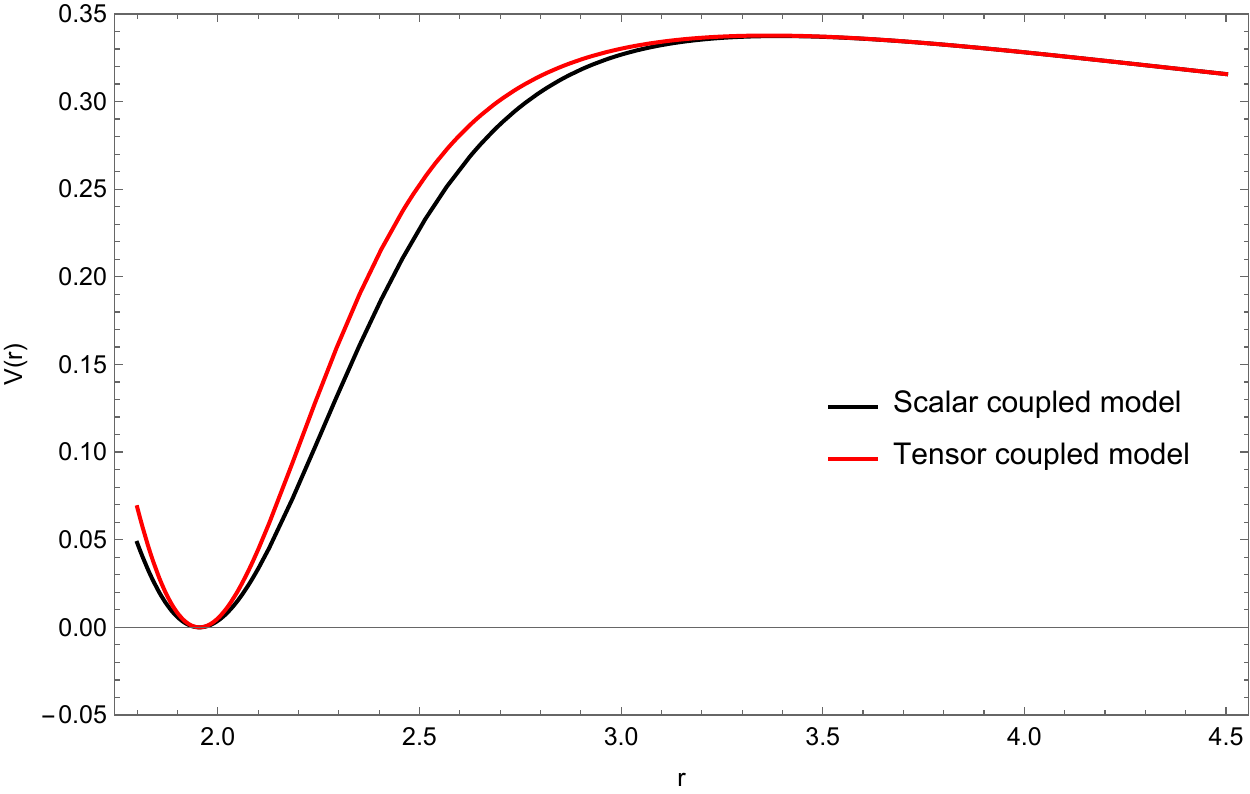}}
\caption{Effective potentials of NC Schwarzschild black holes in two coupling models with $\theta=0.2$, $\ell=2$, and $\mu=0.5$. The thresholds for potential-well formation are  $\zeta_{\text{well}}=6.867$ and $\hat\zeta_{\text{well}}=5.673$.}
\label{f2}
\end{figure}

Although the potential-well formation criterion depicted in Fig.~\ref{f1} is given for the tensor-coupled model, Fig.~(\ref{f2}) demonstrates that the same behavior is also obtained in the scalar-coupled model. Moreover, we have verified that the same condition identifying the onset of potential-well formation is also satisfied for RBHs in NLED which are considered in this paper. This observation motivates us to employ the near-horizon approximation to derive analytical expressions for $\zeta_{\text{well}}$. 

\subsection{Near-horizon analysis and analytical criteria for potential-well formation}

In this subsection, we analyze the behavior of the effective potential in the near-horizon region of static, spherically symmetric black holes. The motivation for this analysis stems from our observation that the formation of an exterior potential well begins when the first local minimum of the effective potential reaches the event horizon. This condition provides a characteristic threshold for the emergence of the potential well, which we denote by $\zeta_{\text{well}}$. 

For RBHs described by the line element (\ref{bm}), the metric function $f(r)$ can be expanded near the event horizon $r=r_h$ as
\begin{equation} \label{ex1}
f(r) \simeq f^{\prime}(r_h)(r-r_h)
+\frac{1}{2}f^{\prime\prime}(r_h)(r-r_h)^2+\cdots,
\end{equation}
where the first term vanishes due to the horizon condition $f(r_h)=0$. By introducing the near-horizon coordinate $x=r-r_h$ and defining the surface gravity as $\kappa=\frac{1}{2}f^{\prime}(r_h)$, the expansion (\ref{ex1}) becomes
\begin{equation} \label{ex2}
f(x)=2\kappa x+\frac{1}{2}f^{\prime\prime}(r_h)x^2+\mathcal{O}(x^3).
\end{equation}
Substituting this expansion into Eqs.~(\ref{veff}) and (\ref{VT}), the effective potentials take the following form in the near-horizon region:
\begin{equation}
V(x)\simeq ax^2+bx,
\end{equation}
where the coefficients $a$ and $b$ for the scalar-coupled model are
\bea \label{p1}
a &\!\!\!\!=\!\!\!\!&
-\frac{8\kappa(\kappa r_h+2(1-\kappa r_h)\zeta+\ell(\ell+1))}{2r_h^3} \notag\\
&\!\!\!\!+\!\!\!\!&
\frac{r_h f''(r_h)\left(6\kappa r_h+(2-24\kappa r_h)\zeta
+\ell(\ell+1)+r_h^2\mu^2-r_h^2\zeta f''(r_h)\right)}
{2r_h^3}, \\
\label{p10}
b &\!\!\!\!=\!\!\!\!&
\frac{2\kappa\left(2\kappa r_h+(2-8\kappa r_h)\zeta
+\ell(\ell+1)+r_h^2\mu^2\right)}{r_h^2}
-2\kappa\zeta f''(r_h),
\eea
while for the tensor-coupled model they are given by
\bea \label{p2}
a &\!\!\!\!=\!\!\!\!&
-\frac{1}{4r_h^2\left(2\hat\zeta r_h\kappa\!-\!\hat\zeta\!+\!r_h^2\right)^2}
\Bigg[r_h f''(r_h)\bigg(\hat\zeta r_h f''(r_h)\Big(6r_h^3\kappa
\!+\!r_h^2(8\hat\zeta\kappa^2\!-\!\ell(\ell+1))\!-\!2\hat\zeta r_h\kappa(\ell(\ell\!+\!1)\!-\!3) \nn\\
&\!\!\!\!+\!\!\!\!&
\hat\zeta\ell(\ell\!+\!1)\Big)\!
+\!2\Big(2r_h^4\kappa(\hat\zeta\mu^2\!-\!3)\!-\!r_h^5\mu^2
\!-\!r_h^3(24\hat\zeta\kappa^2\!-\!\hat\zeta\mu^2\!+\!\ell(\ell\!+\!1))\!+\!\hat\zeta r_h(4\hat\zeta\kappa^2(\ell(\ell\!+\!1)\!+\!7)\!+\!\ell(\ell\!+\!1)) \nn\\
&\!\!\!\!+\!\!\!\!&
6\hat\zeta r_h^2\kappa(3-4\hat\zeta\kappa^2)
+6\hat\zeta^2\kappa(\ell(\ell+1)-2)\Big)\bigg)\!+\! 16\kappa\Bigl(
r_h^4\kappa+r_h^3(\hat\zeta\mu^2+\ell(\ell+1))\nn\\
&\!\!\!\!+\!\!\!\!&
\hat\zeta r_h^2\kappa(-4\hat\zeta\kappa^2+5\ell(\ell+1)-6)+ 2\hat\zeta^2r_h\kappa^2(3\ell(\ell+1)-4)
-\hat\zeta^2\kappa(\ell(\ell+1)-4)\Bigr)
\Bigg], \\
\label{p20}
b &\!\!\!\!=\!\!\!\!&\frac{\kappa\left(\hat\zeta r_h f''(r_h)(\ell(\ell\!+\!1)\!-\!2r_h\kappa)\!+\!2r_h^3\mu^2\!+\!4r_h^2\kappa\!+\!2r_h(4\hat\zeta\kappa^2\!+\!\ell(\ell\!+\!1))
\!+\!4\hat\zeta\kappa(\ell(\ell+1)\!-\!2)\right)} {r_h(2\hat\zeta r_h\kappa-\hat\zeta+r_h^2)}.
\eea

The first derivative of the near-horizon effective potential is $V'(x)=2ax+b$. Then, requiring the extremum to be located at the horizon $x=0$ gives the condition $V'(0)=0$, or equivalently $b=0$. Applying this condition to the explicit near-horizon expansions in two models gives the following analytical expressions for the potential-well formation thresholds:
\begin{equation} \label{crits}
\zeta_{\text{well}} =\frac{r_h^2\mu^2+2r_h\kappa+\ell(\ell+1)}{r_h^2f''(r_h)+8r_h\kappa-2},
\end{equation}
and
\begin{equation} \label{critt}
\hat\zeta_{\text{well}}=\frac{2r_h\bigl[r_h(r_h\mu^2+2\kappa)+\ell(\ell+1)\bigr]}
{\bigl(2r_h\kappa+\ell(\ell+1)\bigr)\bigl(r_hf''(r_h)+4\kappa\bigr)}.
\end{equation}

As illustrated in Fig.~(\ref{f2}), at $\zeta=\zeta_{\text{well}}$ the effective potential vanishes at the horizon $r=r_h$ and its first derivative vanishes as well. Thus, the above expressions can be interpreted as the couplings at which the first extremum of the effective potentials reach the horizon and an exterior potential well begins to form as the coupling increases. One might therefore expect to obtain these expressions directly by solving $V'(r_h)=0$. Indeed, this procedure gives the exact result for the scalar-coupled model. However, for the tensor-coupled model, the more complicated dependence of the effective potential on the Einstein tensor requires the near-horizon expansion to obtain the corresponding analytical expression.

As a consistency check of our analytical criteria for the potential-well formation in both models, we consider the Schwarzschild limit, $f(r)=1-2M/r$. It is well established that the Schwarzschild black hole is linearly stable against scalar perturbations \cite{Wald:1979lth,Dafermos:2016uzj}. In this limit, $\zeta_{\text{well}}\to\infty$ in both models, indicating that no exterior potential well of the type considered here forms for any finite value of the coupling parameter. This is consistent with the absence of the instability mechanism associated with such a well in the Schwarzschild background. Furthermore, Eq.~(\ref{crits}) shows that, in the eikonal limit ($\ell\to\infty$), the Ricci-coupled model has $\zeta_{\text{well}}\to\infty$. As a consequent, for any fixed finite value of the coupling constant, no potential well emerges outside the horizon in this limit.

Moreover, Eqs.~(\ref{crits}) and (\ref{critt}) explicitly show the dependence of the potential-well formation thresholds on the scalar-field mass. For the parameter ranges considered here, increasing $\mu$ raises $\zeta_{\text{well}}$ in both models. This means that a larger coupling is required for the formation of an exterior potential well as the scalar-field mass increases.  On the other hand, in the massless limit $\mu=0$, the threshold in the tensor-coupled model becomes independent of the multipole number $\ell$, i.e.,
\begin{equation} \label{critt0}
\hat\zeta_{\text{well}} =
\frac{2r_h}{r_hf''(r_h)+4\kappa}.
\end{equation}
Interestingly, the same expression is obtained in the eikonal limit ($\ell\to\infty$) of Eq.~(\ref{critt}), even for nonzero scalar-field mass. This explains why the curves for the tensor-coupled model approach a plateau at large $\ell$.

\begin{figure}[H]
\centering
\subfigure [Scalar-coupled model]
{\includegraphics[width=.49\textwidth]{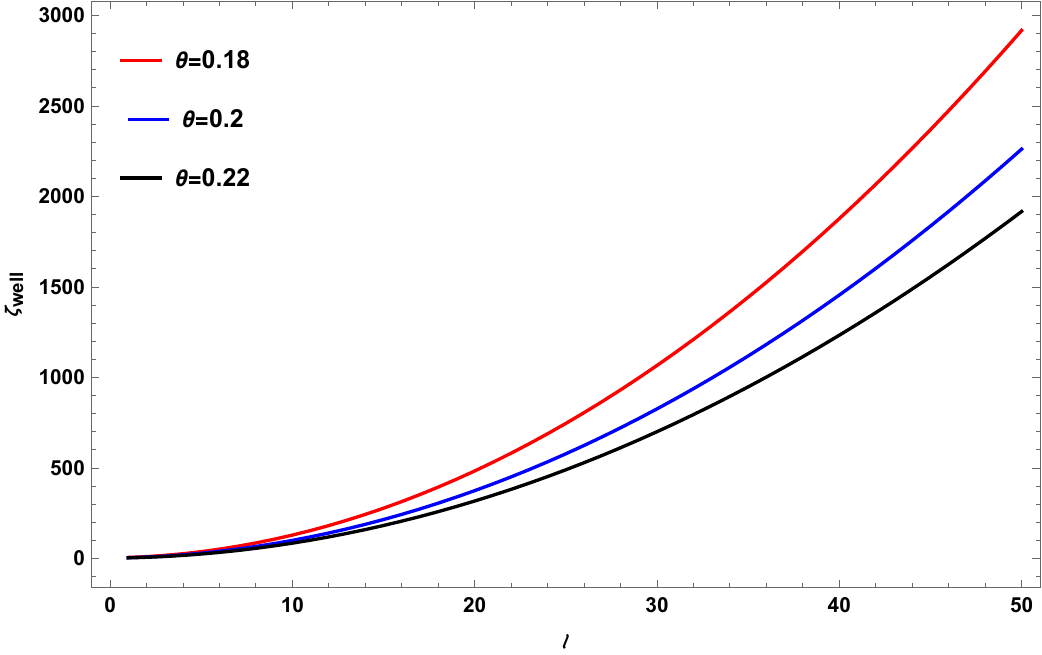}}
\subfigure[Tensor- coupled model]
{\includegraphics[width=.49\textwidth]{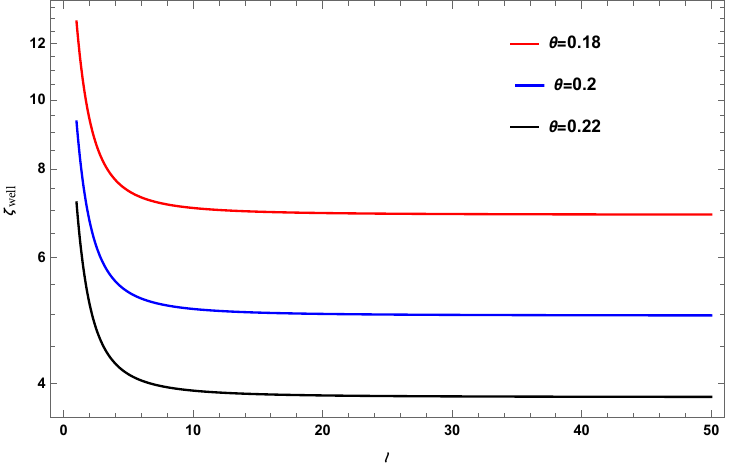}}
\caption{$\zeta_{\text{well}}$ for NC Schwarzschild black holes as a function of $\ell$ when $\mu=0.8$.}
\label{f18}
\end{figure}

\begin{figure}[H]
\centering
{\includegraphics[width=.50\textwidth]{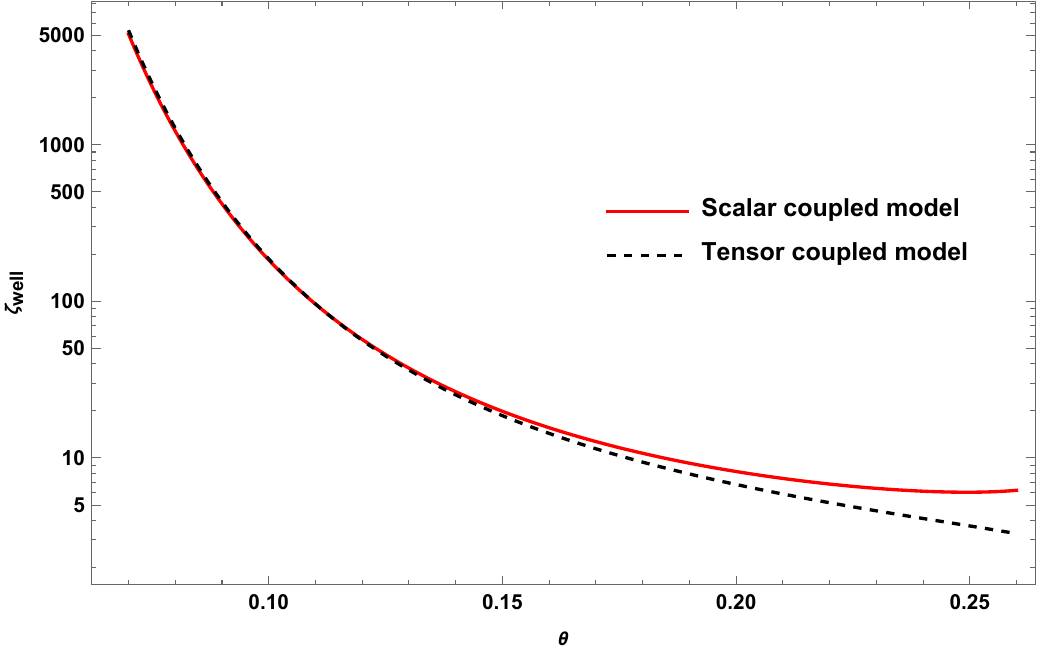}}
\caption{Comparison of $\zeta_{\text{well}}$ in two models as a functions of $\theta$ when $\ell=2$ and $\mu=0.8$.}
\label{f19}
\end{figure}

\begin{figure}[H]
\centering
\subfigure [ $\zeta_{\mathrm{well}}=6.86748$ and $\zeta_{\mathrm{crit}}=12.637$]
{\includegraphics[width=.49\textwidth]{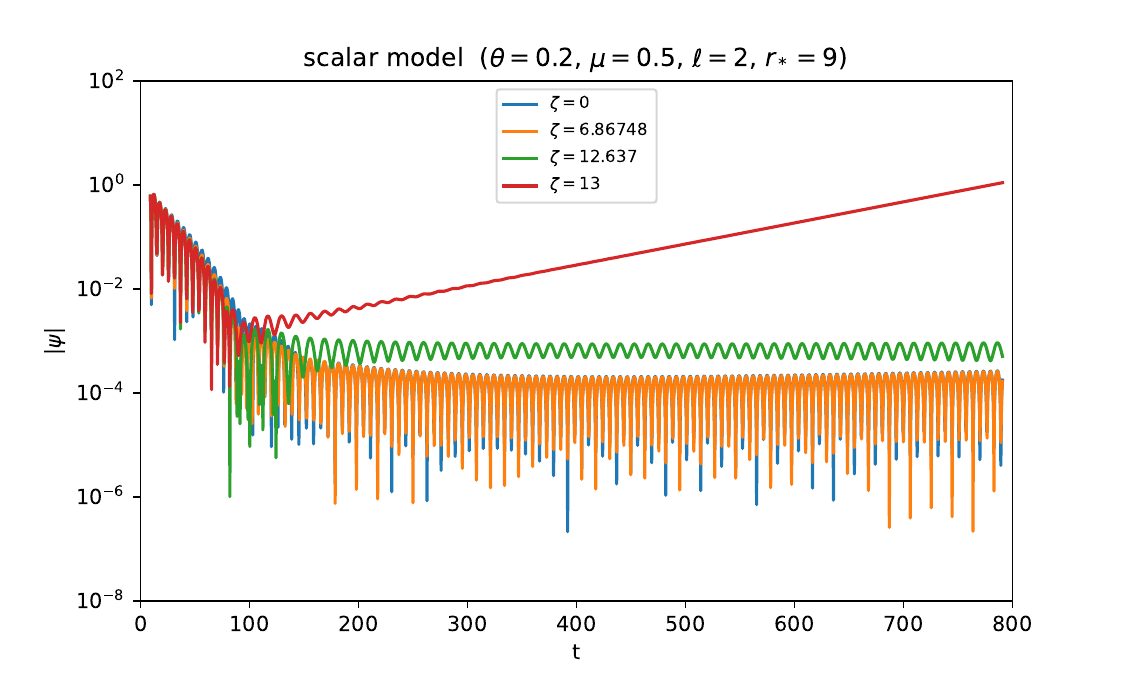}}
\subfigure[$\zeta_{\mathrm{well}}=5.67376$ and $\zeta_{\mathrm{crit}}=14.559$]
{\includegraphics[width=.49\textwidth]{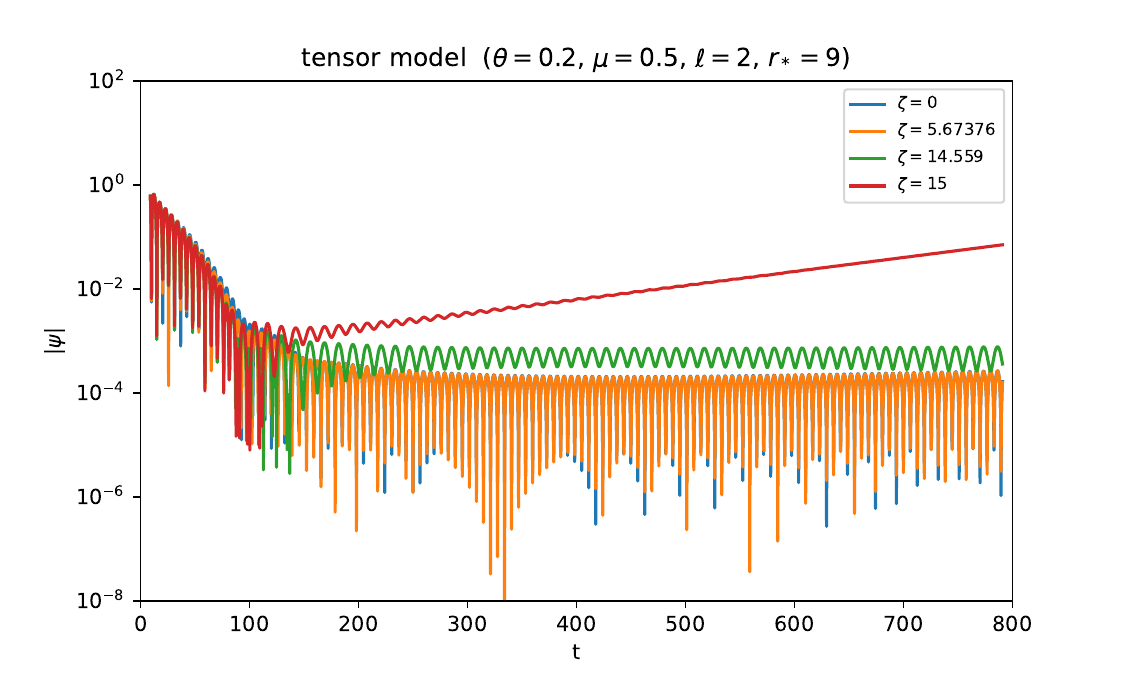}}
\caption{Ringdown profiles for NC Schwarzschild black holes with $M=1$. }
\label{f100}
\end{figure}

\begin{figure}[H]
\centering
\subfigure [Scalar-coupled model]
{\includegraphics[width=.49\textwidth]{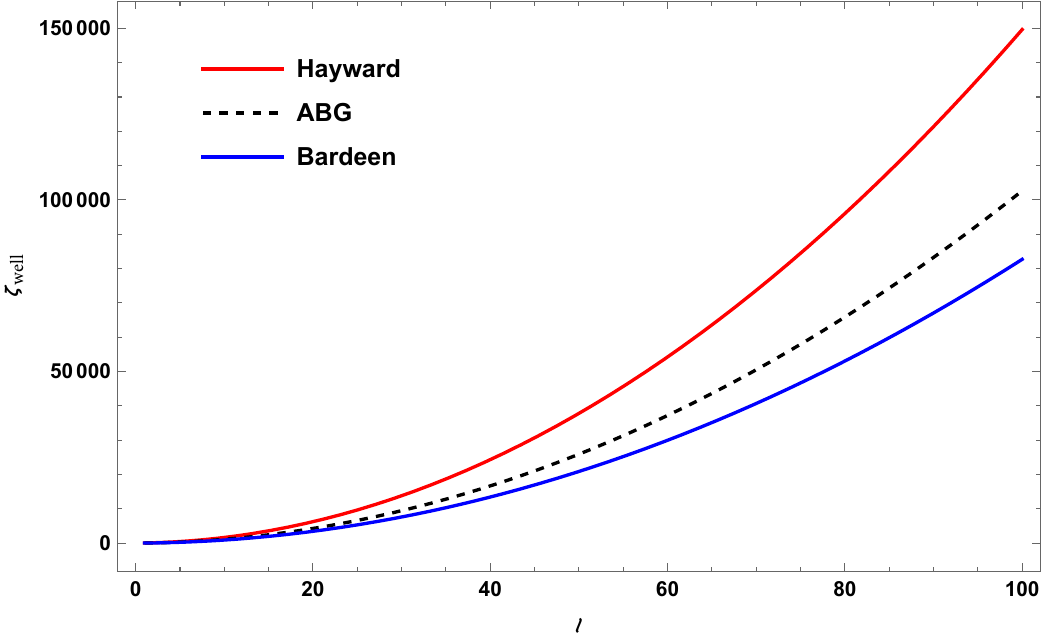}}
\subfigure[Tensor-coupled model]
{\includegraphics[width=.49\textwidth]{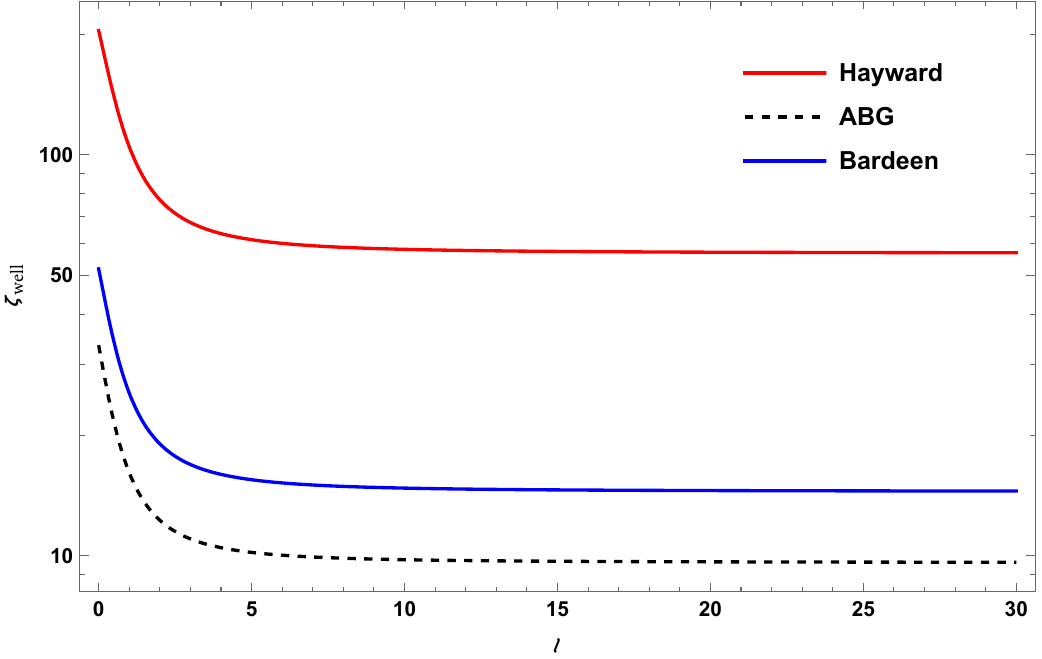}}
\caption{Behavior of $\zeta_{\text{well}}$ for RBHs in NLED theory as a function of $\ell$ with $q=0.45$ and $\mu=0.8$.}
\label{f20}
\end{figure}

\begin{figure}[H]
\centering
{\includegraphics[width=.50\textwidth]{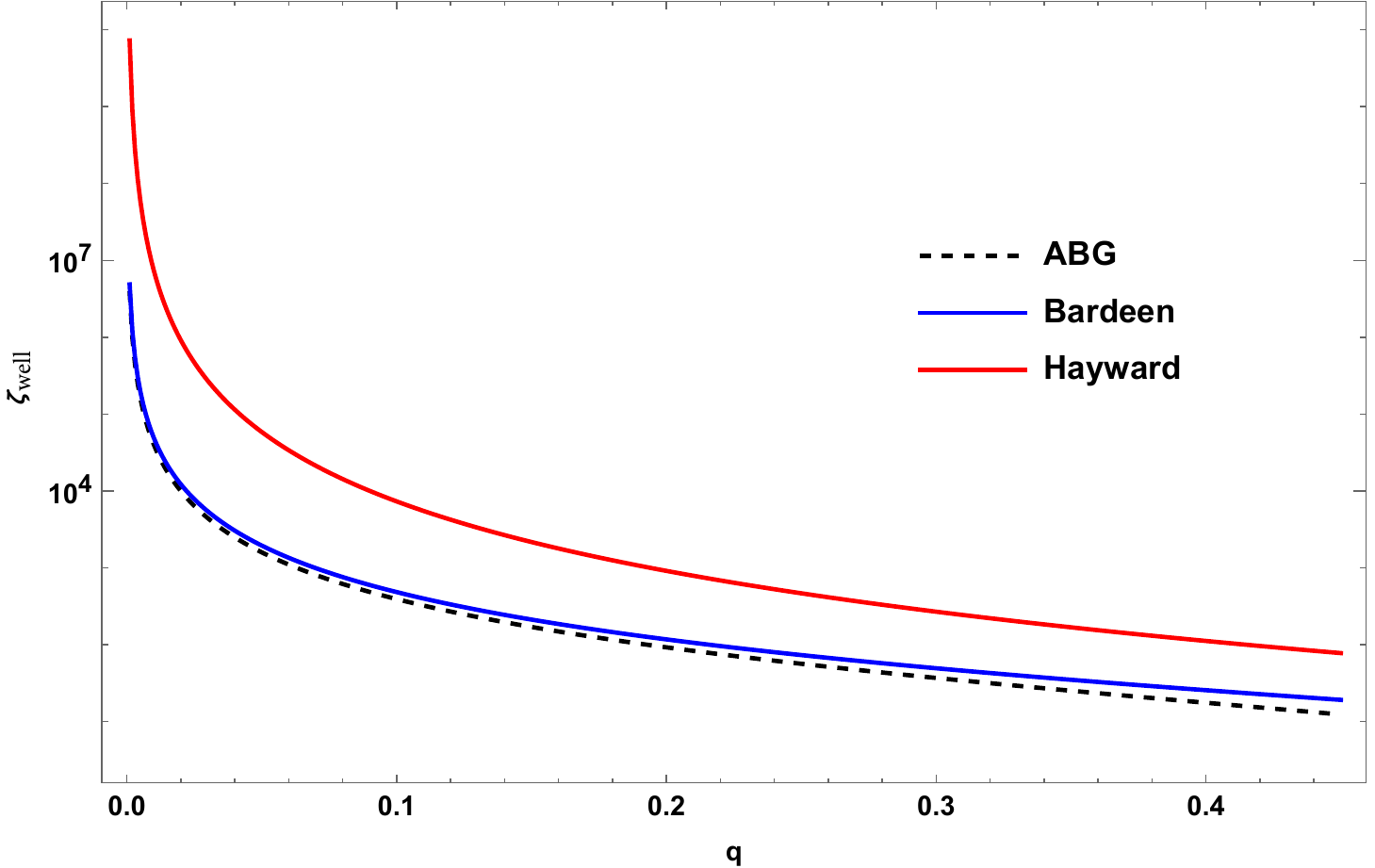}}
\caption{$\zeta_{\text{well}}$ for RBHs in NLED as a function of $q$ in the tensor model with $\ell=2$ and $\mu=0.8$.}
\label{f21}
\end{figure}

\begin{figure}[H]
\centering
{\includegraphics[width=.50\textwidth]{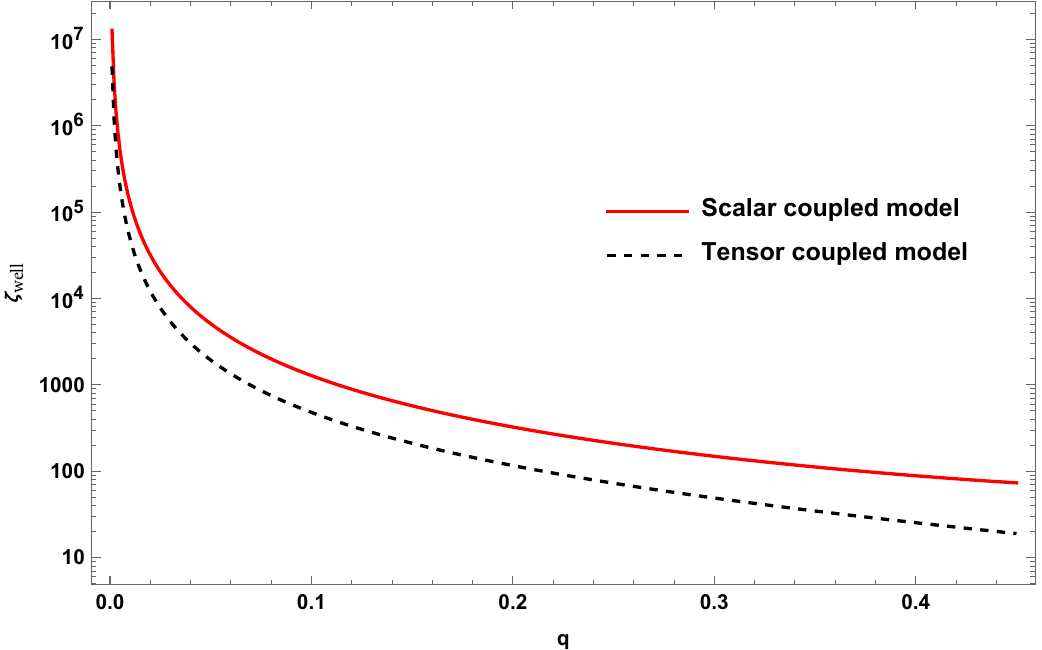}}
\caption{$\zeta_{\text{well}}$ as a function of $q$ for Bardeen RBHs in both coupling models with $\ell=2$ and $\mu=0.8$.}
\label{f22}
\end{figure}

To better understand how the $\zeta_{\text{well}}$ depends on the characteristic parameters of the RBHs considered in this paper, and to compare the two non-minimal coupling models, we plot $\zeta_{\text{well}}$ as a function of $\ell$, $\theta$, and $q$ in Figs.~(\ref{f18})-(\ref{f22}). Since we set $M=1$ in these calculations, similar coupling $\zeta$ can be used in both models for simplicity. For the NC Schwarzschild black holes, Fig.~(\ref{f18}a) shows that $\zeta_{\text{well}}$ increases monotonically with $\ell$ in the scalar-coupled model, while in the tensor-coupled model as shown in Fig.~(\ref{f18}b), it decreases rapidly and eventually approaches a plateau, in agreement with the analytical behavior discussed above in the eikonal limit. It can also be observed from these figures that the potential-well formation threshold decreases as the NC parameter $\theta$ increases which is not explicitly visible in relations (\ref{crits}) and (\ref{critt}). This decreasing trend is particularly clear in Fig.~(\ref{f19}). For small allowed values of $\theta$, the two models yield nearly identical threshold values, whereas a deviation will occur between them at larger values.

It is reasonable to point out that the critical coupling associated with the onset of instability, \(\zeta_{\mathrm{crit}}\), is expected to lie above \(\zeta_{\mathrm{well}}\). Therefore, the increase or decrease of \(\zeta_{\mathrm{well}}\) can also serve as an indicator of the system's tendency toward stability or instability. The implication of instability, here, is that when the imaginary part of QNM becomes positive, the perturbations undergo an exponentially growth and will diverge. This transition from decaying mode to a diverging phase signals an instability which can also be inspected from ringdown profiles. The value of \(\zeta_{\mathrm{crit}}\) corresponds to that point, which has been extensively discussed in our recent work \cite{Karimabadi:2025gpk}. Using the time-domain integration method \cite{Gundlach:1993tn}, we have plotted the ringdown profiles for the NC Schwarzschild black holes in Figs.~(\ref{f100}). 

In order to acknowledge the results in Figs.~(\ref{f18}) and profiles in Figs.~(\ref{f100}) we have provided a numerical comparison for different values of the coupling parameter in Tab.~(\ref{T1}) for the particular value $M/\sqrt{\theta}\simeq2.24$ of authorized values of NC parameter. It is observed that for all values of $\ell$ the critical coupling for onset of instability is larger than the potential-well formation threshold and by increasing $\ell$ these values increase in the scalar model while decrease in the tensor model in matching with curves in Figs.~(\ref{f18}). 
          
\begin{table}[H]
\centering
\caption{$\zeta_{\text{c}}$ and $\zeta_{well}$ for two models when $M/\sqrt{\theta}\simeq2.24$ and $\mu=0.5$}
\label{T1}
\begin{tabular}{c|cc|cc}
\toprule
& \multicolumn{2}{c|}{$\zeta_{\text{c}}$} & \multicolumn{2}{c}{$\zeta_{well}$} \\
\cmidrule(lr){2-3} \cmidrule(lr){4-5}
$\ell$ & scalar model & tensor model & scalar model & tensor model \\
\midrule
2   & 12.637  & 14.559 & 6.86748  &5.67376 \\
4   & 28.123  & 9.052  & 19.2709  & 5.20324  \\
6   & 50.784  & 7.527  & 38.762  & 5.08582  \\
10  & 117.416 & 6.424  & 99.0072 & 5.0177  \\
15  & 240.595 & 5.913  & 214.182 & 4.99455  \\
20  & 408.077 & 5.668  &373.654 & 4.98611  \\
\bottomrule
\end{tabular}
\end{table}   
       
In Fig.~(\ref{f20}), we compare  $\zeta_{\text{well}}$  for three RBHs in NLED as a function of the multipole number $\ell$. In the left panel Fig.~(\ref{f20}a), corresponding to the scalar-coupled model, $\zeta_{\text{well}}$ increases monotonically with $\ell$ for all three backgrounds, following the same trend as that of NC Schwarzschild black holes in Fig.~(\ref{f18}a). Among these geometries, the Hayward black hole exhibits the largest value of $\zeta_{\text{well}}$ throughout the considered range of $\ell$, followed by the ABG and Bardeen black holes, respectively. Thus, within this coupling scheme, a larger coupling is required to form an exterior potential well in the Hayward geometry than the other two backgrounds. In the right panel Fig.~(\ref{f20}b), corresponding to the tensor-coupled model, $\zeta_{\text{well}}$ decreases with $\ell$ and then approaches a plateau for all three backgrounds just like the NC Schwarzschild case. The Hayward black hole again has the largest threshold, but in contrast to the scalar model the Bardeen black hole has a larger threshold than the ABG black hole.

Since RBHs in NLED are charged solutions, it would also be of interest to investigate how the potential-well formation threshold varies with the charge parameter $q$, in correspondence with the $\theta$ consideration for NC Schwarzschild black holes. For instance, we perform this analysis for the tensor-coupled model, as shown in Fig.~(\ref{f21}). The results indicate that $\zeta_{\text{well}}$ decreases rapidly for small values of $q$, while its rate of decrease becomes smaller as $q$ increases. This means that, for the considered backgrounds and parameter ranges, an exterior potential well can form at a weaker coupling as the charge parameter increases. The scalar-coupled model exhibits a similar decreasing trend and the same relative ordering among the considered geometries.

To compare the two coupling schemes directly, we plot $\zeta_{\text{well}}$ as a function of $q$ for the Bardeen RBH in Fig.~(\ref{f22}). As shown, $\zeta_{\text{well}}$ decreases monotonically with increasing $q$ in both models. Moreover, the tensor coupling yields lower potential-well formation thresholds than the scalar coupling throughout the considered range of $q$. It has been verified that the same ordering is also found for the other RBHs in NLED theory.

\subsection{Near-horizon QNM}

In general, QNMs can be classified into three distinct categories: the modes localized near the event horizon, the modes localized near the cosmological horizon, and all-region modes \cite{Burikham:2017gdm}. The latter are associated with the global structure of spacetime and are typically obtained using methods such as the WKB approximation or the asymptotic iteration method applied to the full metric without near-horizon approximations. However, the present analysis focuses on the near-horizon sector, in which the metric function is approximated by its linear expansion around the event horizon. The resulting spectrum should not be identified with the conventional all-region QNM spectrum, but rather interpreted as a distinct class of near-horizon modes governed by the local geometry in the vicinity of event horizon.

The tortoise coordinate introduced in Sec.~(\ref{sec2}) can be integrated in the near-horizon regime to yield
\[
r^* \simeq \frac{1}{2\kappa}\ln x,
\]
so that the wave equation (\ref{sceq}) takes the following form
\begin{equation} \label{eqw}
\frac{d^2 \psi}{d x^2} + \frac{1}{x}\frac{d \psi}{d x}
+ \frac{1}{4 \kappa^2 x^2}(\omega^2 - V(x)) \psi(x) = 0.
\end{equation}
The analytical solution of this equation can be expressed in terms of the confluent hypergeometric and generalized Laguerre functions as
\bea \label{HL1}
\psi(x)=e^{-\frac{\sqrt{a} x}{2 \kappa}} \, x^{\frac{i \omega}{2 \kappa}}\left[c_1 \, U\left(\frac{\frac{b}{\sqrt{a}} \!+\! 2 \kappa \!+\! 2 i \omega}{4 \kappa},
\, \frac{i \omega}{\kappa} \!+\! 1,\, \frac{\sqrt{a} x}{\kappa}\right)+c_2 \, L\left(-\frac{\frac{b}{\sqrt{a}} \!+\! 2 \kappa \!+\! 2 i \omega}{4 \kappa},
\, \frac{i \omega}{\kappa},\, \frac{\sqrt{a} x}{\kappa}\right)\right]
\eea
where \(c_1\) and \(c_2\) are integration constants. To impose the ingoing boundary condition at the event horizon, we examine the behavior of this solution in the limit \(x \to 0\), which gives
\bea \label{HL2}
\psi(x)&\!\!\!=\!\!\!& \frac{c_1 \left( \frac{i}{\kappa} \right)^{-\frac{i \omega}{\kappa}}
(-a)^{-\frac{i \omega}{2 \kappa}}\Gamma\left(\frac{i \omega}{\kappa} \right)
x^{-\frac{i \omega}{2 \kappa}}}{\Gamma \left(\frac{i \left(-b - 2 i \sqrt{-a} \kappa + 2 \omega \sqrt{-a} \right)}{4 \sqrt{-a} \kappa}\right)}+\frac{c_1 \Gamma \left(-\frac{i \omega}{\kappa} \right)x^{\frac{i \omega}{2 \kappa}}}{\Gamma \left(\frac{i \left(-b - 2 i \sqrt{-a} \kappa + 2 \omega \sqrt{-a} \right)}{4 \sqrt{-a} \kappa}- \frac{i \omega}{\kappa}\right)}\nn\\
&\!\!\!+\!\!\!&c_2 \, x^{\frac{i \omega}{2 \kappa}}L\left(-\frac{\frac{b}{\sqrt{a}} + 2 \kappa + 2 i \omega}{4 \kappa},\, \frac{i \omega}{\kappa},\, \frac{\sqrt{a} x}{\kappa}\right).
\eea

The last term represents an outgoing wave; hence, the condition of a purely ingoing solution at the event horizon requires \(c_2=0\). Moreover, the second term also corresponds to an outgoing contribution and must vanish (For more detail about analytic discussion see e.g., Ref.~\cite{MahdavianYekta:2018lnp}). By exploiting the properties of the Gamma function, this condition becomes
\[
\frac{i \left(-2 i \sqrt{-a} \kappa
+ 2 \sqrt{-a} \omega - b\right)}
{4 \sqrt{-a} \kappa}
- \frac{i \omega}{\kappa} = -n,
\]
which yields
\begin{equation}\label{no}
\omega = -\frac{b}{2 \sqrt{-a}}
- i \kappa \left(n + \frac12\right),
\end{equation}
where \(n\) is a non-negative integer. As is evident from Eqs.~(\ref{p1})--(\ref{p20}), the real part of this expression depends on the parameters of the corresponding theory, whereas its imaginary part is determined solely by the background parameter \(\kappa\) and the overtone number \(n\).

As discussed above, the condition for the first extremum of the effective potential to coincide with the event horizon, \(r=r_h\), is \(b=0\). Accordingly, Eq.~(\ref{no}) implies that the near-horizon QNM frequencies become purely imaginary at \(\zeta=\zeta_{\mathrm{well}}\), i.e., their real parts vanish at this threshold. This result characterizes the behavior of the near-horizon sector at the onset of potential-well formation; however, it does not by itself establish the onset of a dynamical instability, which requires a separate spectral analysis of the perturbation equation. Purely imaginary QNM frequencies have also been investigated in recent studies \cite{Ma:2026eaf,Capuano:2026tjy}.


\subsection{Area quantization}

The concept of the area quantization of a black hole has become a fascinating subject since the seminal conjecture proposed by Bekenstein \cite{Bekenstein:1974jk,Bekenstein:1995ju} which the area of a black hole horizon, as an adiabatic invariant, is quantized in units of the Planck length as
\[
A_n=8\pi l_p^2 n,
\]
where \(l_p\) is the Planck length, \(A_n\) denotes the equidistant area spectrum, and \(n\) is a quantum number. It was subsequently proposed \cite{Hod:1998vk} that the spacing of the area spectrum can be understood in terms of the QNM frequencies of a perturbed black hole. Moreover, for a thermodynamic system with energy \(E\), the adiabatic invariant is given by \cite{Kunstatter:2002pj}
\begin{equation}\label{ai}
I=\int\frac{dE}{\Delta\omega(E)}
=\int\frac{T_H dS}{\Delta\omega},
\end{equation}
where \(T_H\) and \(S\) denote the Hawking temperature and entropy, respectively \cite{Hod:1998vk}. According to the Bohr correspondence principle, this adiabatic invariant is quantized as \(I\sim n\hbar\). Thus, for a Schwarzschild black hole with vibrational frequency spacing \(\Delta\omega\), one obtains \cite{Dreyer:2002vy}
\begin{equation}\label{area}
A_n=4\hbar\ln3\cdot n.
\end{equation}

The near-horizon frequencies of black hole perturbations for any spherically symmetric black hole, given by Eq.~(\ref{no}), can be employed to investigate area quantization. In the literature, the highly damped mode approximation is commonly used to determine the black hole area spectrum. As shown in Ref.~\cite{Maggiore:2007nq}, for highly excited modes satisfying \(\omega_I\gg\omega_R\), the vibrational frequency spacing is given by
\[
\Delta\omega=|\omega_I|_n-|\omega_I|_{n-1}.
\]
Using Eq.~(\ref{no}), we consequently obtain
\begin{equation}\label{del}
\Delta\omega=\kappa=2\pi T_H.
\end{equation}
Substituting this result into Eq.~(\ref{ai}) and applying the quantization condition \(I\sim n\hbar\), we obtain the area spectrum
\begin{equation}\label{aq}
A_n=8\pi n,
\end{equation}
where we have used the Bekenstein-Hawking entropy \(S=A/4\) \cite{Bekenstein:1973ur,Hawking:1975vcx}.

On the other hand, as demonstrated in the previous subsection, at \(\zeta=\zeta_{\mathrm{well}}\) the real part of the near-horizon QNM frequency in Eq.~(\ref{no}) vanishes. In this case, the vibrational frequency is determined entirely by the imaginary part, and we again obtain Eq.~(\ref{del}). Thus, at \(\zeta=\zeta_{\mathrm{well}}\), the same area quantization result can be recovered without invoking the highly damped mode approximation. This provides an alternative route to area quantization: the highly damped regime at arbitrary coupling and the purely imaginary near-horizon modes at \(\zeta=\zeta_{\mathrm{well}}\) yield the same vibrational frequency spacing and, consequently, the same area spectrum.

Furthermore, Eqs.~(\ref{del}) and (\ref{aq}) show that the area spectrum obtained either in the highly damped regime or at \(\zeta=\zeta_{\mathrm{well}}\) is independent of the parameters of the specific coupling model, of course it happens when one ignores the back-reaction of the scalar test field. In both cases, the frequency spacing is determined solely by the background surface gravity \(\kappa\), leading to a universal area spectrum.
This result is consistent with \(A_n=8\pi l_p^2 n\) and agrees with the proposal that black holes in Einstein gravity possess an equidistant area spectrum \cite{Kothawala:2008in}.


\section{Conclusion}\label{sec4}

In this work, we investigated potential-well formation in two classes of non-minimally coupled scalar-tensor theories and derived analytical expressions for the threshold $\zeta_{\mathrm{well}}$ for static, spherically symmetric black holes given by Eqs.~(\ref{crits}) and (\ref{critt}). A key result is that the first extremum of the effective potential coincides with the event horizon when the coupling reaches $\zeta_{\mathrm{well}}$. This marks the emergence of an exterior well and provides a necessary, but not sufficient, condition for unstable bound states; the actual instability threshold, $\zeta_{\mathrm{crit}}$, estimated from ringdown profiles, is expected to lie above $\zeta_{\mathrm{well}}$. A comparative analysis was done in Tab.~(\ref{T1}). Thus $\zeta_{\mathrm{well}}$ serves as a qualitative indicator of stability.

The numerical results for two classes of RBHs (NC Schwarzschild black hole and charged black holes in NLED) support these predictions. Increasing the scalar-field mass $\mu$ shifts the threshold toward larger coupling in both models. The multipole number dependence is model dependent: increasing $\ell$ raises $\zeta_{\mathrm{well}}$ in the scalar coupling model but lowers it in the Einstein tensor coupling model, where it approaches a plateau; in the eikonal limit, the threshold diverges in the former but becomes $\ell$-independent and coincides with its massless value in the latter. Increasing the NC parameter $\theta$ or the charge $q$ decreases $\zeta_{\mathrm{well}}$, so an exterior well can form at weaker coupling; among the charged RBHs considered here, the Hayward solution has the largest threshold as well.

Finally, in the near-horizon regime the real part of the QNM frequency vanishes at $\zeta_{\mathrm{well}}$, and the area quantization obtained from highly damped QNMs can be recovered at $\zeta=\zeta_{\mathrm{well}}$ without invoking the highly damped approximation. The resulting area spacing is model independent and depends only on the surface gravity $\kappa$, highlighting the role of $\zeta_{\mathrm{well}}$ in connecting the near-horizon QNM spectrum with the area spectrum.



\end{document}